# Quasicrystals: making invisible materials

All-dielectric photonic quasicrystals may act as zero-refractive-index homogeneous materials despite their lack of translational symmetry and periodicity. Zero-index materials stretch wavelengths to infinity and offer applications in light wave front sculpting and optical cloaking.

Svetlana V. Boriskina

Optical materials that exhibit a vanishing refractive index[1–4] have recently moved from the realm of mathematical curiosity into the practical domain of functional photonic devices. As the refractive index is reduced to zero, both the phase velocity and the wavelength of propagating light waves stretch to infinity. Light can then propagate without any phase accumulation, leaping through the medium as if it were completely invisible for passing photons. Such materials can be created by arranging sub-wavelength constituents into composite lattices, and offer application as directional light sources, innovative lenses and waveguides, insulators, optical cloaks, etc.

To assign an effective index value to a composite metamaterial, it should be possible to treat the material as a homogeneous medium. While homogeneity is often associated with translational invariance, Jian-Wen Dong and colleagues demonstrate[5] that artificial structures that lack translational periodicity yet exhibit long-range order – known as quasicrystals - can behave as zero-index materials. This demonstration opens new opportunities for the development of all-dielectric zero-index materials by extending the concept to a wider class of photonic lattices other than periodic photonic crystals.

Not long ago, quasicrystals themselves were considered a curiosity and deemed impossible, as long-range order – which is a hallmark of crystallinity – was assumed to be inextricably linked to translational periodicity. This view has been shattered by Dan Shechtman's discovery of an 'impossible material' with 10-fold rotational symmetry[6], which spurred a lot of debate, starting with ridicule and ending with the Noble Prize in Chemistry. Quasicrystals discovery was so controversial because local rotational symmetries other than 2-, 3-, 4- and 6-fold ones are incompatible with translational symmetry as illustrated in Fig. 1a,b. Once accepted however, it led to a new definition of a crystal as 'any solid having an essentially discrete diffraction pattern,' such as those shown in Fig. 1d,e for a periodic and a 5-fold quasiperiodic lattice.

Photonic analogs of crystals and quasicrystals[7,8] – artificial arrangements of two or more materials with different refractive indices – now find applications across a wide range of light-based technologies. They are capable of molding the flow of light in many ways, from forbidding the light propagation to enhancing and shaping light emission. Interestingly, owing to their higher rotational symmetries, photonic quasicrystals are more 'homogeneous' than their periodic counterparts for light propagating at different angles. As a result, they allow formation of isotropic and complete bandgaps (for both light polarizations) by using low-index materials, which is not possible with periodic photonic crystals[9]. Furthermore, quasicrystals are not limited by strict constrains on the positions of Bragg peaks in their diffraction patterns, making them well-suited for phase matching for nonlinear optical effects such as frequency conversion[10]. The rich spectrum of optical modes with various degrees of spatial localization also makes quasicrystals useful platforms for multi-mode lasing[11] and multi-color optical sensing[12].

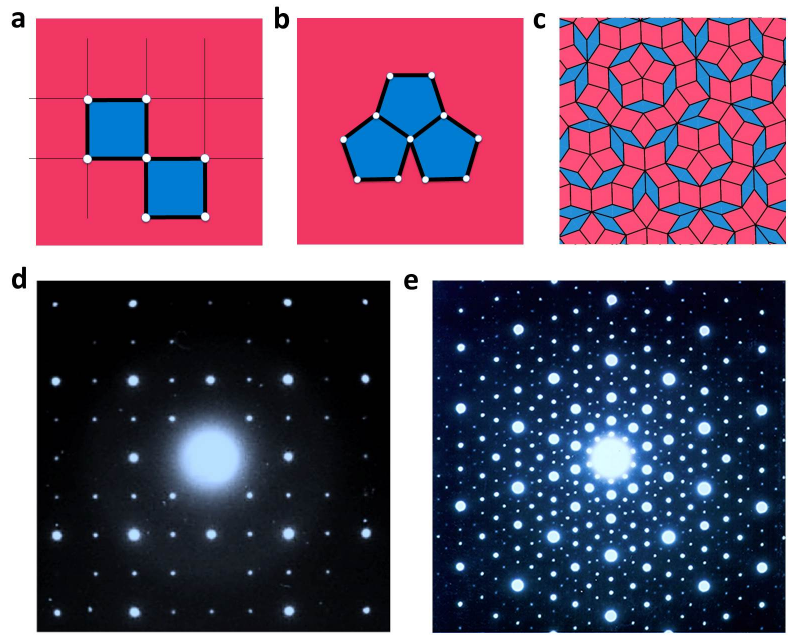

Figure 1| Quasicrystals provide long-range order without structural periodicity. **a**, Periodic lattices with translational periodicities can be formed by tiling unit cells having local rotational symmetries. Only 2-, 3-, 4- and 6-fold symmetries are allowed, such as e.g., in a two-dimensional square-lattice crystal. **b**, 5-, 7-fold and all higher local rotation symmetries are disallowed, as you cannot tile individual unit cells without leaving empty spaces. **c**, Quasicrystals with high-order rotational symmetries – such as the Penrose lattice with 5-fold symmetry – can however be constructed by non-periodic tiling of more than one type of tiles. **d**,**e**, Diffraction patterns of both periodic crystals (**d**) and quasicrystals (**e**) are discrete and feature sharp peaks arranged in periodic lattices for crystals or incommensurate lattices for quasicrystals.

Periodic photonic crystals can act as homogeneous materials, with effective refractive index ranging from positive – typical for conventional materials – to negative, which make materials capable of bending light the 'wrong' way. By tuning their parameters to yield zero effective index, photonic crystals can even be made 'invisible' for light waves[13] (Fig. 2a). Importantly, photonic crystals can act as 'double-zero' materials, with effectively zero permittivity $\varepsilon$ and permeability $\mu$, if they are tuned to exhibit Dirac cone dispersion at the Brillouin zone center

(**k**=0) at a non-zero frequency[13]. Double-zero materials provide impedance matching in addition to the phase invariance achievable in 'single-zero' materials with either $\varepsilon$=0 or $\mu$=0. Dong and co-workers now demonstrate[5] that photonic quasicrystals may also behave as 'double-zero' materials, and this unique property stems from existence of Dirac points at **k**=0 in their dispersion (Figs 2b,c).

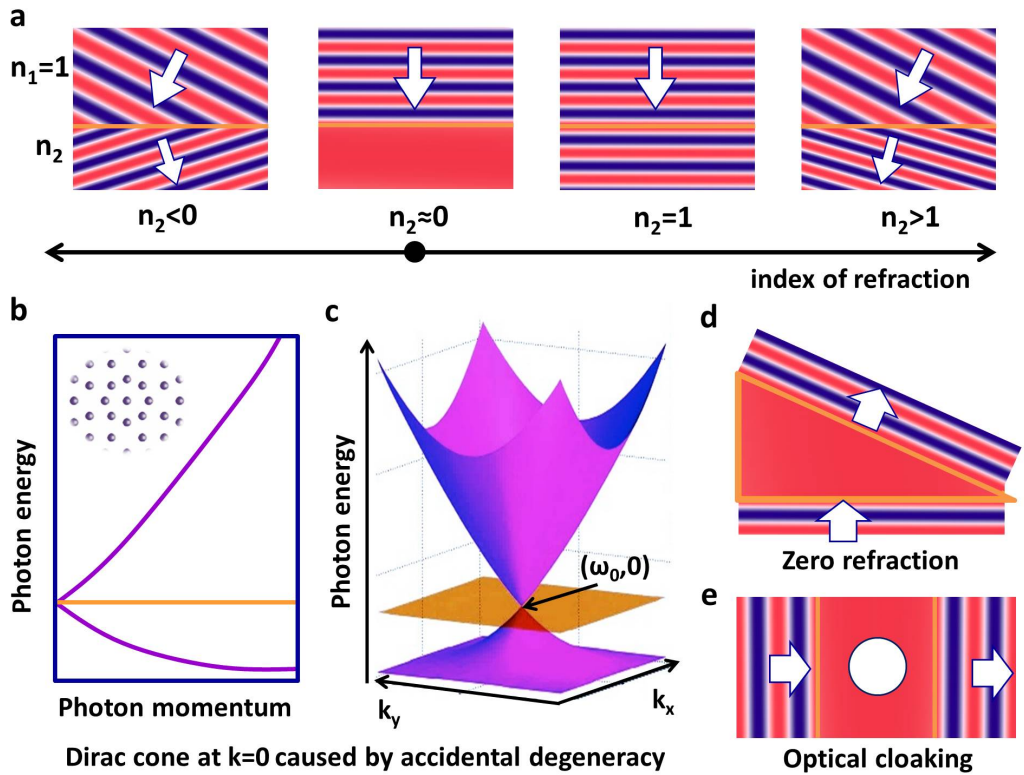

Figure 2| Quasicrystals with Dirac cone dispersion at **k**=0 act as zero effective index materials. **a**, Light waves maintain spatially uniform phase distribution across zero-index material, in striking contrast from both positive- and negative-index media as illustrated for plane wave refraction at the interface. **b,c**, Quasicrystals with high rotational symmetry like the 8-fold lattice shown in the inset provide enough symmetry to protect Dirac cones. Dirac point caused by accidental degeneracy is formed by intersecting bands with linear dispersion as well as an additional flat band. **d**, Plane wave propagating through a prism constructed from a quasicrystal material undergoes zero refraction at interfaces. **e**, 'Invisible' zero-index materials can make embedded objects invisible to light as demonstrated by cloaking of an obstacle inside the quasicrystal developed by *Dong et al*[5]. The images in panels **b**(inset) and **c** are modified from ref. 5, APS.

The relation between the Dirac cone and zero refractive index may not be immediately obvious. However, it is easy to see that if $\varepsilon=\mu=0$ at a non-zero frequency $\omega_0$, the photon dispersion near this frequency is a linear function $\omega=\omega_0+v_g k$ (where $k$ is the wavevector and $v_g$ is the photon group velocity). Linear dispersion is of course a characteristic feature of the Dirac point, which yields many unique transport characteristics of graphene and its photonic analogs. However, the mere existence of the Dirac cone in the bandstructure is not sufficient for the photonic lattice to have zero effective index. Dong *et al*[5] explain that not only a Dirac cone at **k**=0 is required but also it should be formed by crossings of the monopole and dipole branches in the dispersion diagram, and not higher-order multipoles. The latter condition guarantees applicability of the effective medium approximation, while the former ensures that the effective index is local (i.e., does not depend on the wavevector).

The authors provide a recipe for the Dirac cone formation, which is based on tuning the quasicrystal parameters to achieve 'accidental degeneracy' of its modes. This is different from the familiar Dirac cones in graphene etc., which form at the Brillouin zone edge, while nondegenerate bands at the zone center are required by symmetry to be parabolic. The accidental modal degeneracy, which only occurs at certain parameters of the photonic lattice, results in the formation of a Dirac cone accompanied by a flat band that crosses the cone at the Dirac point (Figs. 2b,c).

The same group has previously used this recipe to design and experimentally demonstrate a double-zero material with a periodic photonic crystal lattice[13], and now have extended this concept to a much broader range of materials. The conical dispersions are protected by the lattice symmetry, and this work demonstrates that quasicrystals possess enough rotational symmetry to provide such protection. Furthermore, the authors show that quasicrystal approximants of different sizes feature Dirac cones, albeit at different optimal parameters of the lattice. This is an important finding as quasicrystals of increased size feature higher level of structural complexity and long-range correlations, which sets them apart from their periodic counterparts.

To probe if quasicrystals can act as double-zero effective media, Dong and colleagues designed lattices with 12-fold and 8-fold symmetries for operation at frequencies about 10.5GHz and fabricated them from alumina rods sandwiched between metal plates. The authors showed that the quasicrystal eigenmodes at the Dirac point have almost constant field intensity and phase throughout the whole lattice. They also demonstrated zero refraction of plane waves propagating through a prism constructed from a quasicrystal material (Fig. 2d). Experimental demonstration of the cloaking of an obstacle (a metal rod) inside the quasicrystal (Fig. 2e) was also consistent with the zero index behavior of the material. Finally, another proof that the quasicrystals had zero effective index was delivered by measuring asymmetric wave transport through an inversion-symmetry-breaking prism with a quasicrystal lattice.

Dong et al explain that conditions to achieving zero-index behavior in quasicrystals are more stringent than for opening up photonic bandgaps or even than for realizing negative refractive index. High level of symmetry is required to protect conical dispersion, and while 8-fold and 12-

fold quasicrystals were shown to form Dirac cones, the same feature was not observed in 5-fold Penrose lattices shown in Fig. 1c. Other limitations of this approach include failure of the effective index approximation for low-index structures as well as for vein rather than discrete-rod lattices. Despite these challenges, this research expands the range of geometries for double-zero materials design. It is intriguing if aperiodic structures with long-range order other than quasicrystals[8] can provide symmetry protection required to achieve zero refractive index. Interesting future directions include extension to the visible band and to the third dimension, which may unveil new opportunities for such materials.

*Svetlana V. Boriskina is at the Department of Mechanical Engineering, Massachusetts Institute of Technology, Cambridge, Massachusetts, 02139, USA. e-mail: sborisk@mit.edu*